\documentclass[aps,prl,twocolumn,psfig,groupedaddress,floats,showpacs]{revtex4}

\usepackage{bm}

\usepackage{amssymb,amsmath}
\usepackage{latexsym}
\usepackage{graphicx}
\usepackage{epsfig}

\begin{document}

\preprint{APS/123-QED}

\title{Spin dynamics in (III,Mn)V ferromagnetic semiconductors: 
the role of correlations}

\author{M. D. Kapetanakis and I. E. Perakis}

\affiliation{Department of Physics, University of Crete, 
and 
Institute of Electronic Structure \& Laser, Foundation
for Research and Technology-Hellas, Heraklion, Crete, Greece} 

\date{\today}

\begin{abstract}

We address the role of correlations between  spin and charge 
degrees of freedom  
on the dynamical properties of ferromagnetic 
systems governed by the magnetic exchange interaction 
between itinerant and localized spins.
For this we introduce  a general theory
that treats quantum 
fluctuations beyond the Random 
Phase Approximation based on a correlation expansion 
of the Green's function equations of motion.  
We calculate the spin susceptibility, 
spin--wave excitation spectrum, and 
magnetization precession damping. 
We find that correlations
strongly affect the magnitude and 
carrier concentration dependence of the spin stiffness
and magnetization Gilbert damping.

\end{abstract}

\pacs{75.30.Ds, 75.50.Pp, 78.47.J- }
\maketitle

{\em Introduction---} 
Semiconductors 
displaying  carrier--induced ferromagnetic order, 
 such as  Mn--doped 
III-V semiconductors, manganites, chalcogenides, etc, 
have received a lot of attention due to their
combined magnetic 
and semiconducting properties \cite{rmp,ohno98}.
A strong  response of 
their  magnetic properties to 
carrier density tuning via light, 
electrical gates, or current 
\cite{Kosh,ohno2000,wang07}
can lead to novel spintronics applications \cite{spintron}
and 
multifunctional magnetic devices combining
information processing and  storage
on a single chip.
One of the challenges facing such magnetic  
devices concerns the speed of the basic processing unit, 
determined by  the 
 dynamics of the collective spin.

Two key parameters characterize the spin dynamics in ferromagnets: 
the 
spin stiffness, $D$, and the Gilbert damping 
coefficient, $\alpha$.  
$D$ determines the long--wavelength 
 spin--wave excitation energies, $\omega_{{\bf Q}}\sim D Q^2$,
where ${\bf Q}$ is the momentum, and 
other magnetic properties. 
$D$ also sets an upper limit 
to the ferromagnetic transition temperature:
$T_{c} \propto D$  
\cite{rmp}. 
So far, the $T_{c}$ of (Ga,Mn)As 
has increased 
from $\sim$110 K \cite{ohno98} 
to $\sim$173 K \cite{jung05,rmp}. 
It is important 
for potential  room temperature ferromagnetism 
to consider the theoretical limits of 
$T_c$.

The Gilbert coefficient,
$\alpha$, characterizes the damping of the magnetization 
precession
described by the 
Landau--Lifshitz--Gilbert (LLG) equation \cite{LLG,rmp}.  
A microscopic expression 
can be  obtained by 
relating the  spin susceptibility 
of the LLG equation to the Green's 
function \cite{sinova} 
\begin{equation}
\label{defY}
\ll A\gg = -i
\theta(t) <[A(t), S_{\bf Q}^{-}(0)]>
\end{equation}
with $A=S_{-\bf Q}^{+}$,
 $ S^+=S_x + i S_{y}$.  
$\langle \cdots \rangle$ denotes the average 
over a grand canonical ensemble and ${\bf S}_{{\bf Q}} = 1/\sqrt{N} 
\sum_{j} {\bf S}_{j} e^{- i {\bf Q}{\bf R}_j }
$, where  
${\bf S}_j$ are spins localized at $N$  randomly distributed 
positions ${\bf R}_j$. 
The microscopic origin of $\alpha$ is still not fully understood \cite{sinova}. 
A mean--field calculation of the magnetization damping due to 
the interplay between spin--spin interactions 
and carrier spin dephasing 
was developed in 
Refs.\cite{sinova,green-RPA}.
The  magnetization dynamics can be probed with, e.g.,  
ferromagnetic resonance \cite{ferro}  
and ultrafast magneto--optical pump--probe spectroscopy experiments
\cite{wang07,wang-rev,tolk,chovan}. 
The interpretation of such experiments requires
a better  theoretical understanding 
of dynamical magnetic properties.

In this Letter we discuss the effects of spin--charge correlations, 
due to the p--d exchange coupling of local  and 
itinerant spins,
 on  the 
spin stiffness
and Gilbert damping coefficient.
We describe quantum fluctuations 
beyond the Random Phase Approximation (RPA)
\cite{macdonald-01,konig-01}  with a correlation  expansion \cite{fricke}  
of higher Green's functions 
and a 1/S expansion of the spin self--energy. 
To $O(1/S^2)$, 
we obtain
a strong enhancement,
as compared to the RPA,   
of the spin stiffness 
and the magnetization damping
and a different 
dependence on carrier concentration. 

{\em Equations of motion---} 
The  magnetic properties 
can be described by the Hamiltonian
\cite{rmp}
$H$=$H_{MF}$+$H_{corr}$, where  
the mean field Hamiltonian $H_{MF}$=$\sum_{{\bf k} n}
\varepsilon_{{\bf k}n} a^\dag_{{\bf k} n}a_{{\bf k} n}$
describes 
valence  holes created by 
$a^\dag_{{\bf k} n}$, where
${\bf k}$ is the momentum, $n$  is the band index, 
and $\varepsilon_{{\bf k}n}$  the band dispersion
in the presence of the mean field created by the magnetic exchange interaction
\cite{konig-01}.  
The Mn impurities act as acceptors, creating 
a hole Fermi sea with concentration $c_h$, and 
also provide  $S=5/2$  
local  spins. 
\begin{eqnarray}
\label{Hexch}
H_{\rm{corr}} = \beta c
\sum_{{\bf q}} 
\Delta  S^z_{{\bf q}}  \Delta s^z_{-{\bf q}}
+ \frac{ \beta c}{2} 
\sum_{{\bf q}} 
(\Delta S^+_{{\bf q}} \Delta s^-_{-{\bf q}} 
+ 
h.c.), 
\end{eqnarray}
where 
$\beta$ $\sim$50--150meV nm$^3$ in (III,Mn)V semiconductors \cite{rmp}
is the magnetic exchane interaction. 
$c$ is the Mn spin concentration  
and ${\bf s}_{{\bf q}}=
1/\sqrt{N} 
\sum_{n n^\prime {\bf k}}
{\bf \sigma}_{n n^\prime} 
a^\dag_{{\bf k +q} n}a_{{\bf k} n^\prime}$
the hole spin operator. $\Delta A  = A - \langle A \rangle$
describes the quantum fluctuations of $A$. 
The ground state and thermodynamic properties 
of (III,Mn)V semiconductors in the metallic regime 
($c_h\sim$10$^{20}$cm$^{-3}$) 
are described  to first approximation 
by the 
mean field  virtual crystal approximation, 
$H_{MF}$,
justified for $S \rightarrow \infty$  \cite{rmp}.
Most sensitive 
to the quantum fluctuations induced by  $H_{\rm{corr}}$
are the dynamical properties.
Refs.\cite{macdonald-01,sinova} treated 
quantum effects to $O(1/S)$  
(RPA). 
Here  we study 
correlations that first arise at $O(1/S^2)$.
By choosing the z--axis parallel to 
the ground state local 
spin ${\bf S}$, 
we have   
$S^{\pm} =0$ and $S^z=S$. 
The mean hole spin,  
${\bf s}$, is antiparallel 
to ${\bf S}$, $s^{\pm}=0$ \cite{rmp}.

The spin 
Green's function is given by the equation
\begin{eqnarray}\label{eomY1} 
&&\partial_t 
\ll 
S_{-\bf Q}^{+}\gg 
= -  2 i S \delta(t) + \beta c \ll 
\left({\bf s} \times {\bf S}_{-{\bf Q}}\right)^{+} \gg
 \nonumber \\ 
&&- i \Delta 
\ll s^+_{-{\bf Q}} \gg
+ \frac{\beta c}{N} \times 
 \nonumber \\ 
&&
\sum_{{\bf k} {\bf p} n n^\prime}
\ll \left( {\bf \sigma}_{n n^\prime} \times 
\Delta {\bf S}_{{\bf p} - {\bf k} -{\bf Q}}\right)^+ 
\Delta [a^\dag_{{\bf k}n} a_{{\bf p}n^\prime}]\gg,
\label{eom1} 
\end{eqnarray} 
where 
$\Delta = \beta c S$  
is the mean field spin--flip energy gap 
and ${\bf s} = 1/N \sum_{{\bf k} n} {\bf \sigma_{nn}}f_{{\bf k} n}$ is the 
ground state hole spin.   
$f_{{\bf k} n} 
=\langle 
a^\dag_{{\bf k} n} 
a_{{\bf k} n}\rangle$ 
is the hole population. 
The first line on the right hand side (rhs) 
describes  the mean field precession
of the Mn spin around the mean
hole spin.
The second line on the rhs describes  the RPA 
coupling to the itinerant hole spin
 \cite{green-RPA}, 
while the last line is due to the correlations. 
The hole spin dynamics 
is described by 
\begin{eqnarray} 
\label{X-eom}
&& \left(i \partial_t
-\varepsilon_{{\bf k}n^\prime} +\varepsilon_{{\bf k-Q}n}
\right)
\ll a^\dag_{{\bf k-Q} \uparrow} a_{{\bf k} \downarrow}\gg 
\nonumber \\  
&=& \frac{\beta c}{2\sqrt{N}} \bigg \lbrack \left( f_{{\bf k-Q} n} 
-  f_{{\bf k} n^\prime} \right) 
\ll S^+_{{\bf  - Q}} \gg 
\nonumber \\  
&
+&
\sum_{{\bf q}m} 
\ll  \left({\bf \sigma}_{n^\prime m} \cdot
 \Delta{\bf S}_{{\bf q}}\right) \Delta [a^\dag_{{\bf k-Q}n} 
a_{{\bf k} + {\bf q} m}] \gg \nonumber \\ 
&-&
\sum_{{\bf q}m} 
  \ll  \left(
 {\bf \sigma}_{m n} \cdot \Delta{\bf S}_{{\bf q}}\right) 
\Delta [a^\dag_{{\bf k-Q-q}m} 
a_{{\bf k} n^\prime}] \gg
\bigg \rbrack.
\end{eqnarray}
The first term on the rhs 
gives the RPA contribution \cite{green-RPA}, while  
the last two terms describe correlations. 

The correlation contributions to Eqs.(\ref{eomY1}) and 
(\ref{X-eom}) 
are determined by the 
dynamics of  the interactions  
between a carrier  excitation and  a local spin  fluctuation. This dynamics 
is 
described by 
the Green's functions  
$\ll \Delta {\bf S}_{{\bf p-k-Q}}\Delta [a^\dag_{{\bf k} n} 
a_{{\bf p} n^\prime}] \gg$, whose
 equations of motion 
 couple to  higher Green's functions,
$\ll S a^\dag a a^\dag a \gg$ and 
$\ll S S
 a^\dag a \gg $, describing dynamics of 
{\em three} elementary excitations. 
To truncate the infinite hierarchy, 
we apply a correlation expansion 
\cite{fricke} and 
decompose $\ll S a^\dag a a^\dag  a \gg$ 
into all possible products of the form 
$\langle a^\dag a 
a^\dag a \rangle\ll S\gg$,  
$\langle S\rangle \langle a^\dag a \rangle\ll a^\dag a \gg$, 
$\langle a^\dag a \rangle  
\ll\Delta S \Delta[a^\dag a]\gg$, 
and $\langle S \rangle 
\ll a^\dag a a^\dag a\gg_c$,  
where $\ll a^\dag a a^\dag a\gg_c$
is obtained after subtracting 
all uncorrelated 
contributions, $\langle a^\dag a \rangle 
\ll a^\dag a \gg$, from 
$\ll a^\dag a a^\dag a \gg$
(we include  all  permutations 
of momentum and band indices)  \cite{unpubl}. 
Similarly, we decompose 
$\ll  S  S 
 a^\dag a \gg$ into products 
of the form 
$ \langle S  S \rangle
\ll a^\dag a \gg$,  
$\langle  S \rangle  
\langle  a^\dag a \rangle 
\ll  S \gg$, 
$\langle  S \rangle  
\ll  \Delta S
\Delta [ a^\dag a] \gg$, 
and $\langle  a^\dag a \rangle 
\ll \Delta S \Delta S \gg$.  
This corresponds to decomposing all operators $A$ into average
and quantum fluctuation parts
and neglecting products of three fluctuations.   
We thus describe  all correlations
between any {\em two} spin and charge excitations
and neglect 
correlations among {\em three} or more  
elementary  excitations (which  
contribute to $O(1/S^3)$) \cite{unpubl}.
In the  case of ferromagnetic $\beta$, 
as in the manganites, 
we 
recover the variational results of 
Ref.[\onlinecite{kapet-1}]
and thus 
obtain 
very good 
agreement with exact diagonalization 
results  while reproducing 
exactly solvable limits 
(one electron, half filling, and 
atomic limits, see Refs.[\onlinecite{kapet-1,unpubl}]).

When treating  correlations        in the realistic (III,Mn)V system,  
the numerical solution of the above 
closed system of equations of motion 
is complicated 
by the coupling of 
many momenta and  bands  
and by unsettled issues 
regarding the  role on the 
 dynamical and magnetic anisotropy 
properties of impurity bands, strain, 
localized states, and sp--d hybridization 
 \cite{rmp,burch,wang,sham,anis}.  
In the simpler RPA case, which neglects inelastic 
effects, 
a six--band effective mass approximation \cite{konig-01}  
revealed an order of magnitude enhancement
 of $D$. 
The single--band RPA model 
\cite{macdonald-01}
also predicts maximum $D$ at very small hole concentrations, 
while in the six--band model $D$ increases and then saturates with 
hole doping.
Here we illustrate 
the main qualitative features due to  
ubiquitous correlations important 
in different  ferromagnets
\cite{kapet-1,cmr}  
by adopting the single--band Hamiltonian \cite{macdonald-01}.
We then discuss the role of the multi--band structure of  (III,Mn)V semiconductors
by using a  heavy and  light hole band model.

In the case of two bands of spin--$\uparrow$ and spin--$\downarrow$ 
states \cite{macdonald-01}, we obtain  
by Fourier transformation 
\begin{eqnarray}\label{green-pole} 
\ll S^+_{{\bf -Q}} \gg_{\omega} = 
-\frac{2  S}{
\omega 
+ \delta 
+ \Sigma_{\rm{RPA}}({\bf Q},\omega) +\Sigma_{\rm{corr}}({\bf Q},\omega)},
\end{eqnarray}
where $\delta = \beta c s$ gives the energy splitting of the local spin levels. 
$\Sigma_{\rm{RPA}}$ is the RPA self energy \cite{macdonald-01,konig-01}.  
\begin{eqnarray} 
&&\Sigma_{\rm{corr}}= 
\frac{\beta c}{2N} \sum_{{\bf k p}}
\Bigg \lbrack 
(G_{{\bf p k }\uparrow} + F_{{\bf p k}})
\  \frac{
\omega +\varepsilon_{{\bf k}}-\varepsilon_{{\bf k +Q}}}{\omega+
\varepsilon_{{\bf k}}-\varepsilon_{{\bf k +Q}}+ \Delta + i \Gamma}
\nonumber \\
&& 
-(
G_{{\bf  p k}\downarrow}-
F_{{\bf p k}})
 \frac{\omega+
\varepsilon_{{\bf p-Q}}-\varepsilon_{{\bf p}}
 }{\omega+
\varepsilon_{{\bf p-Q}}-\varepsilon_{{\bf p}}+
\Delta+i \Gamma} \Bigg \rbrack
\label{self-corr} 
\end{eqnarray} 
is the correlated contribution, where 
\begin{equation} 
G_{\sigma}=\frac{\ll S^{+}\Delta[a^\dag_{\sigma} 
a_{\sigma}] \gg}{\ll S^+ \gg} \ , \ 
F=\frac{\ll \Delta S^z a^\dag_{\uparrow} a_{\downarrow} \gg}{\ll S^+\gg}. 
\end{equation}  
 $\Gamma \sim$10-100 meV  is the 
hole spin dephasing rate \cite{Gamma}. 
Similar to 
Ref.[\onlinecite{green-RPA}]
and the Lindblad method calculation of Ref.\cite{chovan},   
we describe such elastic effects by substituting the spin--flip 
excitation energy $\Delta$ by $\Delta + i \Gamma$. 
We obtained $G$ and $F$ by solving the corresponding equations
to lowest order 
in 1/S, with
$\beta S$ kept constant, 
which gives 
$\Sigma^{\rm{corr}}$
to 
$O(1/S^2)$. 
More details will be presented elsewhere \cite{unpubl}.
  
\begin{figure}[t]
\vspace{0.38 in}
\centerline{
\hbox{\psfig{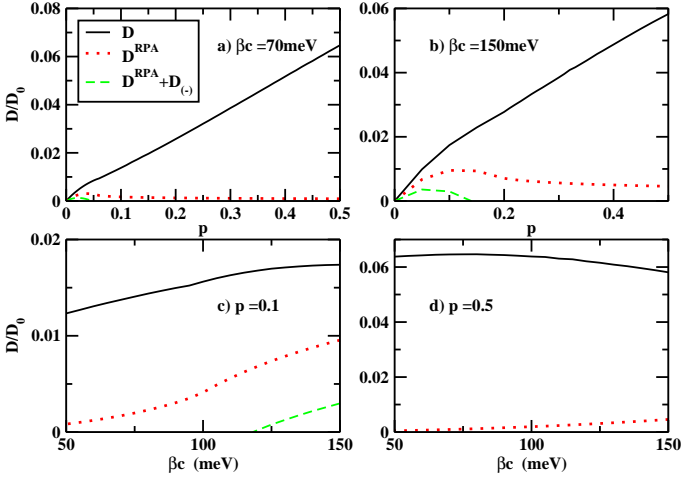}}}
\caption{(Color online) Spin stiffness $D$  
as  function of hole doping
and interaction strength for the 
single--band model. 
$c =1 \rm{nm^{-3}}$, $\Gamma$=0, 
$D_0=\hbar^2/2m_{hh}$, 
$m_{hh}=0.5 m_e$.  
}\label{Fig1}
\end{figure} 

{\em Results---} 
First we study the  spin stiffness 
$D = D^{\rm{RPA}} + D^{\rm{corr}}_{+}
+ D^{\rm{corr}}_{-}$. 
The RPA contribution $D^{\rm{RPA}}$ reproduces 
Ref.[\onlinecite{macdonald-01}].  
The correlated cotributions 
$D^{\rm{corr}}_{+}>0$
and $D^{\rm{corr}}_{-}<0$ 
were obtained to $O(1/S^2)$
from Eq.(\ref{self-corr}) 
 \cite{unpubl}: 
\begin{eqnarray}
D^{\rm{corr}}_{-} 
&=& -\frac{\hbar^2}{2 m_h S^2 N^2}
 \sum_{{\bf k p }} 
 \Bigg \lbrack
\frac{ f_{{\bf k}\downarrow}  (1 - f_{{\bf p} \downarrow})
\  \varepsilon_{{\bf p}} \ ( \hat{{\bf p}} \cdot \hat{{\bf Q}} 
)^2}{\varepsilon_{\bf p}-\varepsilon_{\bf k}} 
\nonumber \\ 
&&+
\frac{f_{{\bf k}\uparrow}  (1 - f_{{\bf p} \uparrow})
\  \varepsilon_{{\bf k}} 
\ ( \hat{{\bf k}} \cdot 
\hat{{\bf Q}} )^2}{
\varepsilon_{{\bf p}}-\varepsilon_{{\bf k}}}
\Bigg \rbrack, \label{D-}
\\
D^{\rm{corr}}_{+} 
&=& \frac{\hbar^2}{2 m_h S^2 N^2}
 \sum_{{\bf k p }} 
f_{{\bf k}\downarrow}  (1 - f_{{\bf p} \uparrow}) \times 
\nonumber \\
&&\left[
\varepsilon_{{\bf k}} 
\ ( \hat{{\bf k}} \cdot 
\hat{{\bf Q}} )^2
+ \varepsilon_{{\bf p}} 
\ ( \hat{{\bf p}} \cdot 
\hat{{\bf Q}} )^2
\right]  \times 
\nonumber \\ 
&&\left[
\frac{2}{\varepsilon_{{\bf p}}-\varepsilon_{{\bf k}}}
+ \frac{1}{\varepsilon_{{\bf p}}-\varepsilon_{{\bf k}} + \Delta}
-\frac{\Delta}{(\varepsilon_{{\bf p}}-\varepsilon_{{\bf k}})^2} \right],
\label{D+}
 \end{eqnarray} 
where ${\bf \hat{Q}}$, ${\bf \hat{k}}$, and 
${\bf \hat{p}}$ 
denote the unit vectors.

For ferromagnetic 
interaction, as in the manganites \cite{cmr,kapet-1}, 
the Mn and carrier spins align in parallel.
The  Hartree--Fock 
is then the  state of maximum  spin 
and  an exact eigenstate 
of the many--body Hamiltonian (Nagaoka state).  
For anti--ferromagnetic 
$\beta$, as in  (III,Mn)V semiconductors, 
the ground state carrier spin is anti--parallel to the  Mn spin
and can increase via the scattering 
of a spin--$\downarrow$  hole 
to an empty spin--$\uparrow$
state (which decreases $S_z$  by 1). 
 Such quantum fluctuations 
give rise to 
$D^{\rm{corr}}_{+}$, 
Eq.(\ref{D+}),  which vanishes for $f_{{\bf k}\downarrow}=0$. 
$D^{\rm{corr}}_{-}$ comes from magnon  scattering 
accompanied by the creation of 
a Fermi sea pair.  
In the case of  a spin--$\uparrow$ Fermi sea, 
Eq.(\ref{D-}) recovers the 
results of Refs.\cite{kapet-1,cmr}.

\begin{figure}[t]
\vspace{0.38 in}
\centerline{
\hbox{\psfig{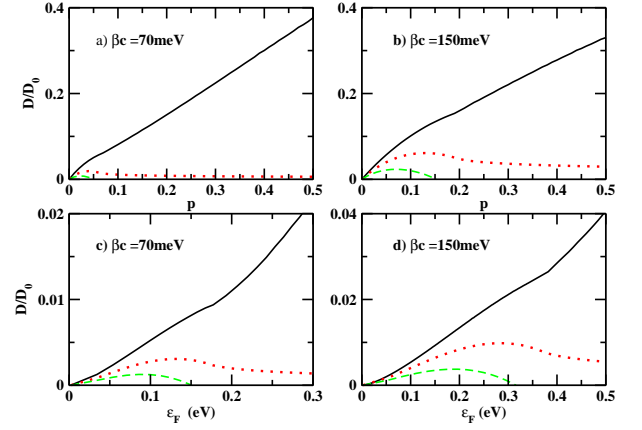}}}
\caption{(Color online) Spin  stiffness $D$ for
the parameters of Fig. \ref{Fig1}.  
(a) and (b):  two--band model, (c) and (d): 
dependence on the Fermi energy  
within the single--band model.
} \label{Fig2}
\end{figure} 

We evaluated Eqs.(\ref{D-}) and (\ref{D+})
for zero temperature 
after introducing an upper energy cutoff 
corresponding to the 
Debye momentum, $k_D^3=6 \pi^2 c$, that ensures  the correct number 
of magnetic ion degrees of freedom \cite{macdonald-01}. 
Figs. \ref{Fig1}(a) and (b)   show the 
dependence of $D$ on  hole doping, 
characterized by $p = c_h/c$, 
for two couplings $\beta$, 
while 
Figs. \ref{Fig1}(c) and (d) 
show its dependence on  $\beta$ 
for two dopings $p$.
Figure \ref{Fig1} 
also compares our full result, $D$, with $D^{\rm{RPA}}$
and  $D^{\rm{RPA}}+D^{\rm{corr}}_{-}$.
It is clear that the correlations beyond RPA
have a pronounced effect on the spin stiffness, 
and therefore on $T_c \propto D$ \cite{jung05,rmp} 
and other magnetic properties. 
Similar to the manganites \cite{cmr,kapet-1}, 
$D^{\rm{corr}}_{-}<0$ 
 destabilizes the ferromagnetic phase.
However, $D^{\rm{corr}}_+$
 strongly enhances $D$
 as compared to $D^{\rm{RPA}}$ \cite{macdonald-01} 
and also changes its  p--dependence.

The ferromagnetic order and $T_c$ values observed 
in (III,Mn)V semiconductors 
cannot be explained 
with the single--band 
RPA approximation \cite{macdonald-01},  
which predicts a  small  $D$ 
that 
decreases with increasing $p$. 
Figure \ref{Fig1}  
shows that the correlations  change these RPA results in a profound way. 
Even within the single--band model, the correlations 
strongly  enhance $D$ and change its 
$p$--dependence: $D$ now 
increases with $p$.
Within the RPA, such behavior can be obtained only 
by including 
multiple valence bands 
\cite{konig-01}.
As discussed e.g. in Refs.\cite{jung05,rmp},   
the main bandstructure effects can be 
understood by considering two  bands
of heavy ($m_{hh}$=0.5$m_e$) and light (
$m_{lh}$=0.086$m_e$) holes.
$D$ is dominated and enhanced by the  more dispersive light hole band. 
On the other hand, the
heavily populated heavy 
hole states dominate the static properties and $E_F$. 
By adopting such a two--band model, 
we obtain the results of Figs. \ref{Fig2}(a) and (b).
The main difference from Fig. \ref{Fig1}  
is the order of magnitude enhancement of all contributions,  
due to $m_{lh}/m_{hh}=0.17$. 
Importantly, the  differences between  $D$ and $D^{\rm{RPA}}$ 
remain strong. Regarding the upper limit of $T_c$
due to collective effects,  
we note from Ref.\cite{jung05} 
that is is proportional to $D$ and the mean field Mn spin.  
We thus expect an enhancement, as compared to the RPA result, 
comparable to the difference between $D$ and $D^{\rm{RPA}}$.

The doping dependence of $D$ mainly comes from its 
dependence on  $E_F$, 
shown in Figs. \ref{Fig2}(c) and (d), 
which differs strongly from the RPA result. 
Even though the two band model captures 
these differences,
it fails to describe accurately the dependence of $E_F$ on $p$, 
determined by the successive  population of multiple 
anisotropic bands.  
Furthermore, the spin--orbit interaction reduces
the hole spin matrix elements 
\cite{sham}.  
For example, $|\sigma^{+}_{nn^\prime}|^2$ 
is maximum when the band states 
are also spin eigenstates. The spin--orbit interaction mixes 
the spin--$\uparrow$ and spin--$\downarrow$ states 
and reduces $|\sigma^{+}_{nn^\prime}|^2$. 
From Eq.(\ref{eomY1}) 
we see that the deviations from the mean field result 
are determined by the coupling to the Green's functions  $
\ll \sigma_{nn^\prime}^+\Delta [a^\dag_{n} a_{n^\prime}]\gg$ (RPA), 
$\ll \Delta  S^z
\sigma_{nn^\prime}^+ 
\Delta [a^\dag_{n} a_{n^\prime}]\gg$ 
(correction to RPA due to $S^z$  fluctuations  
leading to 
$D^{\rm{corr}}_{+}>0$), 
and 
$\ll \Delta S^+
\sigma_{nn^\prime}^z 
\Delta [a^\dag_{n} a_{n^\prime}]\gg$
(magnon--Fermi sea pair scattering leading to  
$D^{\rm{corr}}_{-}<0$). 
Both the RPA and the correlation contribution
arising from $\Delta S^z$
are proportional to 
$\sigma_{nn^\prime}^+$. Our main result,
i.e. the {\em relative} importance of the correlation as compared to the 
RPA contribution,
should thus also hold in the realistic system. 
The full solution 
will be pursued elsewhere.

\begin{figure}[t]
\vspace{0.38 in}
\centerline{
\hbox{\psfig{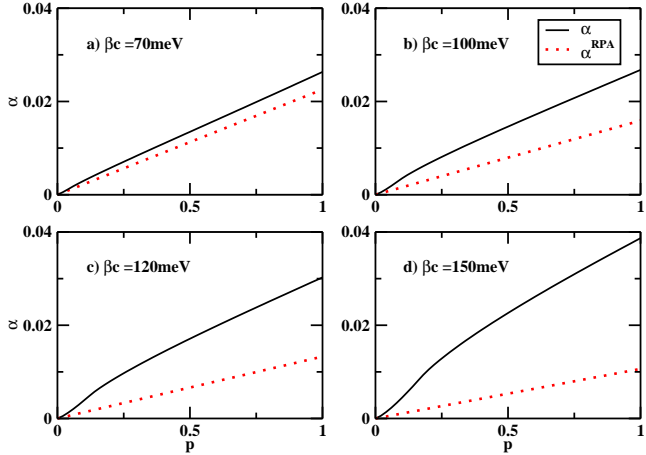}}}
\caption{(Color online) Gilbert damping 
as  function of hole doping for different interactions $\beta$.
$c =1 \rm{nm}^{-3}, \Gamma =20{\rm meV}$. 
}\label{Fig3}
\end{figure}

We now turn to the Gilbert damping coefficient, 
$\alpha=2S/\omega \times \rm{Im} \ll S^+_{{\bf 0}} \gg^{-1}$     
at $\omega \rightarrow 0$ \cite{sinova}.
We obtain to $O(1/S^2)$ that $\alpha = \alpha^{\rm{RPA}} + 
\alpha^{\rm{corr}}$, 
where 
$\alpha^{\rm {RPA}}$
recovers the mean--field 
result of Refs \cite{green-RPA,sinova}
and predicts a {\em linear} dependence on the hole doping $p$, while  
\begin{eqnarray}
&&\alpha^{\rm{corr}} = 
\frac{\Delta^2}{2N^2S^2} \sum_{{\bf k p }} 
\rm{Im} \Bigg \lbrack
\frac{f_{{\bf k} \downarrow} ( 1 - f_{{\bf p} \uparrow} )}{
\Delta + i \Gamma} 
 \times \nonumber \\
&& 
\left(\frac{1}{
\varepsilon_{\bf p}-\varepsilon_{\bf k} - \delta } 
+ \frac{1}{
\varepsilon_{\bf p}
-\varepsilon_{\bf k}+  \Delta + i \Gamma} \right) \Bigg \rbrack
\label{a-Born}
\end{eqnarray}
arises from the carrier spin--flip quantum fluctuations.
Fig.(\ref{Fig3})  compares $\alpha$ with the RPA result 
 as  function
of $p$.
The correlations enhance  $\alpha$ and lead to a 
{\em nonlinear} dependence on  $p$, 
which suggests the possibility of controlling the 
magnetization relaxation by tuning 
the hole density.   
A nonlinear dependence of $\alpha$ on  photoexcitation 
intensity was reported in Ref.\cite{tolk} (see also 
Refs.\cite{wang-rev,wang}).  

We conclude that spin--charge correlations 
play an important role on the dynamical 
properties of ferromagnetic semiconductors. 
For quantitative statements, they must be addressed 
together with the bandstructure effects particular 
to the individual systems.
The correlations  studied here should play an important role in the ultrafast 
magnetization dynamics 
observed with pump--probe magneto--optical  spectroscopy
\cite{wang-rev,tolk,chovan,wang,sham}. 
This work was supported by the EU STREP program HYSWITCH.


\begin{references}

\bibitem{rmp} 
T. Jungwirth {\em et al.}, Rev. Mod. Phys. {\bf 78}, 2006.


\bibitem{ohno98}
H. Ohno, Science {\bf 281}, 951 (1998).


\bibitem{Kosh}
S. Koshihara
{\it et al}.,
 Phys. Rev. Lett. {\bf 78},
4617 (1997).
 
\bibitem{ohno2000} H. Ohno {\it et al}., Nature {\bf 408}, 944 (2000).

\bibitem{wang07}
J. Wang {\it et al}.,
Phys. Rev. Lett. {\bf 98},
217401 (2007).

\bibitem{spintron} 
S. A. Wolf {\it et al}., Science {\bf 294}, 1488 (2001).


\bibitem{jung05} 
T. K. Jungwirth {\it et al}., Phys. Rev. B {\bf 72}, 
165204 (2005). 


\bibitem{LLG}
L. D. Landau, E. M. Lifshitz, and L. P. Pitaeviski,
Statistical Physics, Part 2 (Pergamon, Oxford, 1980).


\bibitem{sinova}
J. Sinova {\em et. al.},
Phys. Rev. {\bf B} 69, 085209 (2004); 
Y. Tserkovnyak, G. A. Fiete, and B. I. Halperin,
Appl. Phys. Lett. {\bf 84}, 25 (2004).



\bibitem{green-RPA}
B. Heinrich, D. Fraitov\'a, and V. Kambersk\'y,
Phys. Stat. Sol. {\bf 23}, 501 (1967).


\bibitem{ferro} 
S. T. B. Goennenwein {\it et al}., Appl. Phys. Lett. 
{\bf 82}, 730 (2003).

\bibitem{wang-rev} 
J. Wang {\it et al}.,
 J. Phys: Cond. Matt. {\bf 18},
R501 (2006).


\bibitem{tolk} 
J. Qi {\it et al}.,
Appl. Phys. Lett. {\bf 91}, 112506 (2007).


\bibitem{chovan} 
J. Chovan, E. G. Kavousanaki, 
 and I. E. Perakis, Phys. Rev. Lett. 
 {\bf 96}, 057402 (2006); 
J. Chovan and I. E. Perakis, Phys. Rev. B {\bf 77}, 
085321 (2008).


\bibitem{macdonald-01}
J. K\"{o}nig, H--H Lin and A. H. MacDonald,
Phys. Rev. Lett. {\bf 84}, 5628, (2000); 
M. Berciu and R. N. Bhatt, 
Phys. Rev. B {\bf 66}, 085207 (2002). 

\bibitem{konig-01} 
J. K\"{o}nig, T. Jungwirth, and A. H. MacDonald, 
Phys. Rev. B {\bf 64}, 184423 (2001).

\bibitem{fricke} 
J. Fricke, Ann. Phys. {\bf 252}, 479 (1996). 

\bibitem{unpubl} 
M. D. Kapetanakis and I. E. Perakis, arXiv:0806.0938v1.

\bibitem{kapet-1}
M. D. Kapetanakis, A. Manousaki, and I. E. Perakis,
Phys. Rev. B {\bf 73}, 174424 (2006); 
M. D. Kapetanakis and I. E. Perakis,
Phys. Rev. B {\bf 75}, 140401(R) (2007). 

\bibitem{burch} 
K. S. Burch 
{\em et. al.}, Phys. Rev. Lett. {\bf 97}, 087208 (2006). 

\bibitem{wang}
J. Wang {\em et. al.}, arXiv:0804.3456; 
K. S. Burch {\em at. al.}, 
Phys. Rev. B 70, 205208 (2004).

\bibitem{sham} 
L. Cywi\'{n}ski and L. J. Sham, Phys. Rev. B 
{\bf 76}, 045205 (2007). 

\bibitem{anis}  
X. Liu {\em et. al.}, Phys. Rev. B {\bf 71}, 
035307 (2005); K. Hamaya {\em et. al.}, Phys. Rev. B 
{\bf 74}, 045201 (2006). 

\bibitem{cmr} 
D. I. Golosov, Phys. Rev. Lett. {\bf 84}, 3974 (2000); 
N. Shannon and A. V. Chubukov,
Phys. Rev. B {\bf 65}, 104418 (2002). 

\bibitem{Gamma} T. Jungwirth {\em et. al.}, 
Appl. Phys. Lett. {\bf 81}, 4029 (2002). 

 

\end{references}
\end{document}